\documentclass[11pt]{article}

\usepackage{amsmath,amssymb,amsthm,amsfonts,latexsym,fullpage,listings,graphicx,setspace, multirow, hhline, eurosym, mathrsfs}
\usepackage{mathtools, nccmath} 
\usepackage{bbm} 
\usepackage{caption} 
\usepackage{bm} 

\usepackage{titlesec}

\usepackage{appendix} 
\usepackage{natbib} 
\setcitestyle{authoryear}

\usepackage{chngcntr}
\counterwithin{figure}{section}
\counterwithin{table}{section}

\usepackage{subcaption}

\usepackage{hyperref}
\hypersetup{hidelinks}

\newcommand{\R}{\mathbb{R}}
\newcommand{\ep}{\epsilon}
\newcommand{\de}{\delta}

\usepackage{acronym}
\newacro{ad}[A-D]{\textit{modified Anderson-Darling test}}
\newacro{aic}[AIC]{\textit{Akaike's Information Criterion}}
\newacro{ama}[AMA]{\textit{advanced measurement approach}}
\newacro{b2}[Basel II]{\textit{the second Basel accords}}
\newacro{bcbs}[BCBS]{\textit{Basel Committee on Banking Supervision}}
\newacro{bic}[BIC]{\textit{Bayesian Information Criterion}}
\newacro{bur}[BUR]{\textit{Burr distribution}}
\newacro{cdf}[CDF]{\textit{cumulative distribution function}}
\newacro{evt}[EVT]{\textit{extreme value theory}}
\newacro{gnh}[GNH]{\textit{$g$-and-$h$ distribution}}
\newacro{gpd}[GPD]{\textit{generalized Pareto distribution}}
\newacro{lda}[LDA]{\textit{loss distribution approach}}
\newacro{ldce}[LDCE]{\textit{Loss Data Collection Exercise}}
\newacro{lgn}[LGN]{\textit{lognormal distribution}}
\newacro{lgngpd}[LGNGPD]{\textit{lognormal body spliced with generalized Pareto tail}}
\newacro{lgnlgn}[LGNLGN]{\textit{lognormal body spliced with lognormal tail}}
\newacro{llog}[LLOG]{\textit{Loglogistic distribution}}
\newacro{lsas}[LSAS]{\textit{log-Sinh-arcSinh distribution}}
\newacro{mle}[MLE]{\textit{maximum likelihood estimation}}
\newacro{oos}[OOS]{\textit{Out-of-Sample}}
\newacro{or}[OR]{\textit{operational risk}}
\newacro{orc}[ORC]{\textit{operational risk category}}
\newacro{orx}[ORX]{\textit{ORX Global Database}}
\newacro{pdf}[PDF]{\textit{probability density function}}
\newacro{pmle}[PMLE]{\textit{penalized maximum likelihood estimation}}
\newacro{qs}[QS]{\textit{quantile scoring function}}
\newacro{rc}[RC]{\textit{regulatory capital}}
\newacro{rrc}[RRC]{\textit{regulatory risk category}}
\newacro{rv}[RV]{regularly varying}
\newacro{sas}[SAS]{\textit{Sinh-arcSinh distribution}}
\newacro{subx}[SUBEX]{subexponential}
\newacro{supx}[SUPEX]{superexponential}
\newacro{wbl}[WBL]{\textit{Weibull distribution}}

\title{\bf On the Selection of Loss Severity Distributions to Model Operational Risk}
\author{
	\begin{tabular}{ccc}
	Daniel Hadley & Harry Joe & Natalia Nolde \\
	University of British Columbia & University of British Columbia & University of British Columbia \\
	daniel.hadley@stat.ubc.ca & harry.joe@stat.ubc.ca & natalia@stat.ubc.ca \\
	\end{tabular}}

\date{October 18, 2018}

\begin{document}
\maketitle

\begin{abstract}
Accurate modeling of operational risk is important for a bank and the finance industry as a whole to prepare for potentially catastrophic losses.  One approach to modeling operational is the loss distribution approach, which requires a bank to group operational losses into risk categories and select a loss frequency and severity distribution for each category.  This approach estimates the annual operational loss distribution, and a bank must set aside capital, called regulatory capital, equal to the 99.9\% quantile of this estimated distribution.  In practice, this approach may produce unstable regulatory capital calculations from year-to-year as selected loss severity distribution families change.  This paper presents truncation probability estimates for loss severity data and a consistent quantile scoring function on annual loss data as useful severity distribution selection criteria that may lead to more stable regulatory capital.  Additionally, the Sinh-arcSinh distribution is another flexible candidate family for modeling loss severities that can be easily estimated using the maximum likelihood approach.  Finally, we recommend that loss frequencies below the minimum reporting threshold be collected so that loss severity data can be treated as censored data.
\end{abstract}


\section{Introduction} \label{S:Intro}

The \acf{bcbs} defines \ac{or} in \ac{b2} as
\begin{quote}
\textit{``the risk of a loss resulting from inadequate or failed internal processes, people and systems, or from external events.  This definition includes legal risk but excludes strategic and reputational risk."} \citep*{BCBS2006}
\end{quote}
To mitigate potential operational losses, regulators require that a bank set aside a minimum amount of capital for one year called \ac{rc}.  

The \ac{ama} presented in \ac{b2} calculates \ac{rc} as the $99.9\%$ quantile of the estimated firm-wide annual operational loss distribution.  This distribution is commonly estimated using the \acf{lda}, which requires a bank to estimate and select the loss frequency and loss severity distributions for each \ac{rrc} and treat firm-wide annual operational loss as a compound sum of the annual losses from each \ac{rrc}.

A major obstacle facing banks using the \acs{ama} is to develop a procedure that provides stable \ac{rc} calculations year-to-year.  Large increases in \ac{rc} that result from poor modeling procedures require the bank to set aside more assets that cannot be used freely to generate revenue, so a bank has an intrinsic interest in minimizing the probability of overestimating \ac{rc}.  Regulators scrutinize large decreases in a bank's \ac{rc}, so a bank may also face problems in subsequent years when the overestimation is corrected.  Stability in \ac{rc} is a common problem faced by industry that is not well covered in the operational risk literature.  

Loss frequencies for a given \ac{rrc} tend to follow predictable patterns, so the instability in \ac{rc} is primarily generated by the loss severity distribution selection process.  A popular selection process uses a list of candidate severity distribution families, estimates each distribution using historical operational loss data, and selects the ``best" distribution based on some criteria.  When a \ac{rrc}'s selected severity distribution changes families from one year to the next, that \ac{rrc}'s contribution to firm-wide annual operational loss is likely to change dramatically.  For this reason, we are interested in finding a flexible distribution family that can outperform other candidates year-after-year.

Operational loss data usually exhibit extreme positive-skew and can have various right-tail behaviors.  The log transform of operational loss severities, called log-losses, also exhibit positive skew.  As a result, candidate distribution families should be able to fit loss or log-loss data that is asymmetric with various tail behaviors.  Two-parameter distribution families, while popular in industry and easy to estimate, are not flexible enough to accomplish both goals.  Spliced and mixture distributions are flexible enough to fit asymmetric, right-skewed data, but require the estimation of additional parameters for splicing points or component proportions.  The four-parameter \ac{gnh} {\citep{Hoaglin1985, Dutta2006, Degen2007}} is a flexible distribution family sometimes used to model loss severities, but presents challenges when estimated by \ac{mle}.  The \ac{sas} \mbox{\citep{Jones2009}} provides us with a flexible four-parameter candidate distribution family that is able to accurately fit log-loss severity data for many \ac{rrc}'s.  By treating loss severities as the exponential transformation of a \ac{sas} random variable, we introduce the \ac{lsas} as a viable loss severity candidate distribution family.  \ac{lsas} maintains the flexibility of \ac{gnh}, but is easier work with numerically.

For each \ac{rrc}, we estimate each candidate severity distribution family using historical data of operational losses from that \ac{rrc}. Historical loss data are truncated, so the data are comprised of only those losses exceeding a known minimum reporting threshold.  For each candidate severity distribution family, \ac{mle} is used to estimate the parameters.  Finally, one severity distribution is selected from the estimated candidate distributions.  This paper is primarily focused with the selection process.

Severity distributions are commonly evaluated by performance measures inspired by the five listed in \citet{Dutta2006}.  Hypothesis tests for appropriate severity distributions include Kolmogorov-Smirnov and Anderson-Darling.  Through simulation, we found that these tests have low power and tend to include too many inappropriate distributions.  Relative measures of fit may include \ac{aic} \citep{Akaike1974} and \ac{bic} \citep{Schwarz1978}, however neither criterion can tell us if any distribution is a good fit.  \citet{Dutta2006} consider a severity distribution ``realistic" if it produces regulator capital no more than 3\% of a bank's assets.  However, this criterion is applied to \ac{rc} and cannot be immediately applied to each \ac{rrc}.

We consider two additional severity distribution selection criteria that can be applied to each \ac{rrc}, the truncation probability estimate and the \ac{qs} \citep{Gneiting2011}.  Details for our selection criteria are given in Section \ref{SS:SDSC}.  When estimating a distribution family from left-truncated data with a known truncation point, evaluating the \ac{cdf} at the truncation point given the estimated parameters produces an estimate of the proportion of losses that occur below this threshold.  Thus, we can consider a candidate severity distribution ``realistic" if its truncation probability estimate is reasonable.  We take a rather conservative stance and consider any truncation probability estimate less than 50\% to be reasonable.  This criterion can be applied to each \ac{rrc} independently, making it particularly useful for the \ac{lda}.  To better understand the range of reasonable truncation probability estimates, we promote the collection of loss frequencies below the truncation point.  We present improved truncation probability estimates when using censored data in Section \ref{SS:cens} to motivate the collection of all loss frequencies.

The second criterion, the \ac{qs}, uses a consistent scoring function to evaluate forecasts.  Since \ac{b2} defines \ac{rc} as the forecasted 99.9\% quantile of the estimated firm-wide annual loss distribution, the performance of an \ac{or} modeling procedure should be measured by its ability to predict the extreme right-tail of this distribution.  Under the assumption of weak dependence between the annual losses of each \ac{rrc}, the sum of the 99.9\% quantile of each \ac{rrc}'s estimated annual loss distribution can be used as a proxy for \ac{rc}.  Thus, we can evaluate the \ac{qs} on the annual losses for each \ac{rrc} and select the severity distribution with the best (lowest) score.  The \ac{qs} criterion is used in our simulation study in Section \ref{SS:QSAnn} to rank loss severity distributions and consistently favors appropriate distributions.

The struggle of financial institutions to produce stable \acs{rc} calculations could be a contributing factor in the decision by \ac{bcbs} to move away from the \acs{ama}.  In the next manifestation of the Basel accords, Basel III, the \acs{ama} is being removed from the regulatory framework.

\begin{quote}
\textit{``The option to use an internal model-based approach for measuring operational risk - the `Advanced Measurement Approaches' (AMA) - has been removed from the operational risk framework. BCBS believes that modeling of operational risk for regulatory capital purposes is unduly complex and that the AMA has resulted in excessive variability in risk-weighted assets and insufficient levels of capital for some banks."} 
\citep{BCBS2016}
\end{quote}

Despite this proposed change in regulations, a bank still has plenty of motivation to use the \ac{ama} for internal modeling of their \acs{or}.  According to the article, ``The Final Bill - financial crime" \citep*{Economist2016}, there have been 188 settlements from 2009 to August 2016 for criminal and civil prosecutions against banks costing \$219 billion.  Eleven firms have paid penalties in excess of 10\% of their market capitalization.  In March 2018, Barclays was ordered to pay \$2 billion in civil penalties for fraudulently selling mortgage securities that contributed to the 2008 financial crisis and in April 2018, Wells Fargo was fined \$1 billion for the ``bank's failures to catch and prevent problems, including improper charges to consumers in its mortgage and auto-lending businesses." (Strasburg, 2018; Hayashi, 2018).

In Section \ref{S:LDA}, we briefly review the \ac{lda}.  Our severity distribution estimation and selection process is presented in Section \ref{S:Severity}, while Section \ref{S:Freq} illustrates the importance of the truncation probability estimate on the loss frequency and its role in estimating the annual loss distribution.  A simulation study is performed in Section \ref{S:Sim} that portrays the instability of \ac{rc} when selecting a severity distribution based on \ac{aic} and the modified Anderson-Darling test, compares the \ac{gnh} and \ac{lsas} distribution families, examines improvements to truncation probability estimation when using censored data, and gauges the performance of selecting the severity distribution from the \ac{qs} on annual loss data.  All simulations and numerical analyses use the $\texttt{R}$ software available at https://www.r-project.org/.  Section \ref{S:Conc} summarizes our conclusions and suggests areas for future research.

\section{Loss Distribution Approach} \label{S:LDA}

Under the \acs{b2} \acs{ama} guidelines, operational loss events are partitioned into eight business lines and seven event types.  When an operational loss occurs, it is mapped to one of 56 business line/event type intersections, called a \acf{rrc}.  This mapping process is further explained in \citet{BCBS2006}.  Each loss event must be assigned a time stamp, a loss amount, and a \ac{rrc}.  The time stamp is usually a fiscal quarter or year and the loss amount is a positive value.  The number of loss events that occur in a particular \ac{rrc} over a given time period is called the loss frequency, and each loss amount is called the \textit{loss severity}.

Historical loss data are used to estimate the loss frequency and severity distributions by interpreting the historical loss data as realizations of random variables.  Under an \ac{ama}, the historical loss data should reflect all current, material activities, risk exposures, and loss events whose loss severities exceed a minimum threshold.  For example, a bank that sells off or discontinues a business line should no longer use those losses in their \ac{ama}.  Using notation adapted from \citet{Embrechts2011}, loss events are denoted as
\begin{equation}
\big\{ X_{t, n}^{b, l}\big\}, \quad \text{for $t = 1, 2, ..., T$; $b = 1, 2, ..., 8$; $l = 1, 2, ..., 7$; $n = 1, 2, ..., N_t^{b, l}$},
\label{E:lossrrc}
\end{equation}
where $X_{t, n}^{b, l}$ is a random variable for the loss severity of the $n^{\text{th}}$ loss event occurring in fiscal year $t$ for business line $b$ and event type $l$, and $N_t^{b,l}$ is a random variable for the number of losses occurring in fiscal year $t$ for business line $b$ and event type $l$.  Thus, $S_{T+1}$ is a random variable for firm-wide annual operational loss for next fiscal year and can be calculated as
\begin{equation}
S_{T+1} = \sum_{b=1}^8 \sum_{l=1}^7 \sum_{n=1}^{N^{b,l}_{T+1}} X_{T+1, n}^{b, l} = \sum_{b=1}^8 \sum_{l=1}^7 L^{b, l}_{T+1},
\label{E:totalrrc}
\end{equation}
where $L^{b,l}_{T+1}$ is the annual loss for business line $b$ and event type $l$ in year $T + 1$.  The goal of the \ac{lda} is to estimate the distribution of $S_{T+1}$ via simulation and calculate \ac{rc} as the 99.9\% quantile of this estimated distribution.

There are two sources of randomness in $S_{T+1}$, the loss frequency and the loss severity.  The loss frequency, $N_{T+1}^{b,l}$, is a discrete random variable for the number of losses occurring next fiscal year in business line $b$ and event type $l$.  The loss severity, $X_{T+1,n}^{b,l}$, is a non-negative, continuous random variable as defined in \eqref{E:lossrrc}.  When calculating \acs{rc}, we follow two common assumptions:
\begin{itemize}
  \item{$N_{t}^{b,l}$, are independent of $X_{t,n}^{b,l}$ for a given business line $b$, event type $l$, and year $t$;}
  \item{$\{X_{t, n}^{b,l}\}_{t=1, 2, \dots; \ n=1, 2, \dots, N_t^{b,l}}$ are independent and identically distributed for a given business line $b$ and event type $l$.}
\end{itemize}

For the purpose of estimating loss severity distributions, losses mapped to a given event type may be combined across business lines to produce an \ac{orc}, which is the level at which the bank's model generates a separate distribution to estimate potential losses \citep{BCBS2011}.  We assume an annual loss frequency with no trend.  A general solution to modeling trends in the loss frequency is presented in \citet{CD2015}.  These simplifying assumptions help us avoid complexities that do not contribute to the understanding of loss severity distribution selection.  Since \ac{rc} is an annual forecast, the use of an annual loss frequency mitigates both structural reporting bias and temporal clustering of losses that were evidenced by the 2004 \acf{ldce} \citep{Dutta2006}.  By assuming no trend, we can assume annual loss frequencies are independent and identically distributed for each \ac{orc}.  Finally, working with \ac{orc} loss frequencies allows us to disregard how frequencies are combined across \ac{rrc}'s and simplifies our notation.  Letting $r$ represent a specific \ac{orc}, loss events can be rewritten from Equation \eqref{E:lossrrc} as
\begin{equation*}
\big\{ X_{t, n}^{r}\big\}, \quad \text{for $t = 1, 2, ..., T$; $r = 1, 2, ..., R$; $n = 1, 2, ..., N_t^{r}$},
\label{E:lossorc}
\end{equation*}
and the firm-wide annual operational loss from Equation \eqref{E:totalrrc} becomes
\begin{equation}
S_{T+1} = \sum_{r=1}^R \sum_{n=1}^{N^{r}_{T+1}} X_{T+1, n}^{r} = \sum_{r=1}^R L^{r}_{T+1}.
\label{E:totalorc}
\end{equation}

When internal data is too sparse for an \acs{orc}, \acs{b2} allows a banks' internal data to be supplemented by an external database.  Some external datasets include losses from banks of various sizes located all over the world.  Hence, care should be taken to filter the data so that they are appropriate in size and scope.  Additionally, an external database may have a minimum reporting threshold for loss data collection that differs from a bank's.  If external data are used to supplement internal data, the higher of the bank's and database's reporting threshold may need to be applied to all data for that \acs{orc}.  Since some \acs{orc}'s may be supplemented by external data and others may not, a bank can have different minimum reporting thresholds for different \acs{orc}'s.

\section{Loss Severity Estimation and Selection} \label{S:Severity}

Regulations require that a bank's internal operational loss data include all loss events whose severities exceed a minimum threshold, so we assume that we have no information about the loss events that occur below the threshold.  As a result, the datasets are treated as left-truncated.  In Section \ref{SS:cens}, we investigate the impact of data collection for the loss frequency below the reporting threshold by comparing truncation probability estimates to censored probability estimates.  The remainder of this section includes an overview of the truncation and censoring approaches, presents candidate severity distributions, and introduces our severity distribution selection criteria.

\subsection{Truncation Approach} \label{SS:Trunc}
The truncation approach assumes that losses below the minimum threshold belong to the same distribution as the losses above the threshold.  In order to use the \ac{mle} approach on truncated datasets to estimate the loss severity distributions, we must derive the likelihood function from the conditional density given the truncation point.  For $n$ loss events exceeding the minimum threshold in a given \ac{orc}, assume loss amounts $X_1, X_2, ..., X_n \ \stackrel{iid}{\mathrm{\sim}} \ F$, where $F$ is some loss severity distribution with parameters $\bm{\theta} \in \bm{\Theta}$.  If we let $\tau$ represent the non-random minimum reporting threshold of the given \ac{orc}, then given the truncation point $tau$, the conditional \ac{cdf} for a reported loss is
\begin{equation*}
\widetilde{F}(x; \bm{\theta}, \tau) = \frac{F(x; \bm{\theta}) - F(\tau; \bm{\theta})}{1 - F(\tau; \bm{\theta})}.
\label{E:CDF}
\end{equation*}
The conditional \acf{pdf} for a reported loss is
\begin{equation*}
\tilde{f}(x; \bm{\theta}, \tau) = \frac{d}{dx}\widetilde{F}(x; \bm{\theta}, \tau) = \frac{f(x; \bm{\theta})}{1 - F(\tau; \bm{\theta})}.
\label{E:PDF}
\end{equation*}

We can estimate $\bm{\theta}$ via \ac{mle} on a sample, $\mathbf{x}$, using the likelihood function
\begin{align}
\widetilde{L}(\bm{\theta}; \mathbf{x}, \tau) = \prod_{i=1}^{n} \ \tilde{f}(x_i; \bm{\theta}, \tau) = \big[ 1 - F(\tau; \bm{\theta}) \big]^{-n} \prod_{i=1}^n f(x_i; \bm{\theta}),
\label{condL}
\end{align}
by maximizing $\widetilde{L}(\bm{\theta}; \mathbf{x}, \tau)$ over all $\bm{\theta} \in  \bm{\Theta}$. The \ac{mle} is denoted $\widehat{\bm{\theta}}$.

The truncation probability is the probability that a loss event occurs but does not exceed the minimum threshold and is estimated by $F(\tau; \widehat{\bm{\theta}})$.  When estimating the annual loss distribution, we simulate losses from the unconditional severity distribution to account for the downward bias of treating a truncated sample as complete \mbox{\citep{Baud2003, Chernobai2005, Luo2007}}.

\subsection{Censoring Approach} \label{SS:Cens}
When estimating the candidate distribution families using the truncation approach, each distribution has a truncation probability estimate.  When loss severity data are generated by a process with \ac{rv} tail behavior, the truncation probability estimates produced by distribution families with \ac{subx} tail behavior can become unreasonably ($>50\%$) large \citep{Perline2005}.  Without data collection or expert opinion for the frequency of losses below a reporting threshold, however, it can be challenging to decide a reasonable range of values for the truncation probability.  To encourage banks to start collecting the frequency of losses below the reporting threshold, we compare the truncation probabilities estimated from datasets truncated at $\tau$ to the censored probabilities estimated from the same datasets censored at $\tau$.  The results of our simulation, presented in Section \ref{SS:cens}, make a compelling argument that this additional data collection may be worthwhile.

Similar to the truncation approach, the censoring approach also assumes losses above and below the reporting threshold follow the same distribution.  To derive the likelihood function, assume loss severities are a sequence of random variables, $X_1, X_2, ..., X_n \ \stackrel{iid}{\mathrm{\sim}} \ F$, for some loss severity distribution $F$ with parameters $\bm{\theta} \in \bm{\Theta}$.  Let $\tau$ represent the non-random minimum reporting threshold.  Then, let $Y_i = \max(\tau, X_i)$ and $Y_{i_1}, Y_{i_2}, \dots, Y_{i_m} > \tau$ with $0 \leq m \leq n$, so that we have full loss severity data for $m$ of the $n$ loss events.  The likelihood function for censored data is
\begin{align}
\widetilde{L}_c(\bm{\theta}; \mathbf{x}, \tau) = \big[ F(\tau; \bm{\theta}) \big]^{n-m} \ \prod_{i\in I_m} f(x_i; \bm{\theta}),
\label{censL}
\end{align} where $I_m = \{i = 1, 2, \dots, n : X_i > \tau \}$.
By maximizing $\widetilde{L}_c(\bm{\theta}; \mathbf{x}, \tau)$ over all $\bm{\theta} \in  \bm{\Theta}$, we find the \ac{mle}, $\widehat{\bm{\theta}}_c$.  The censoring probability estimate is calculated as $F(\tau; \widehat{\bm{\theta}}_c)$.

\subsection{Candidate Distributions} \label{SS:Dist}
For each \ac{orc}, we assume that loss severities are generated from one of the following parametric distributions: \ac{lgn}, \ac{gpd}, \ac{bur}, \ac{wbl}, \ac{llog}, \ac{gnh}, \ac{lsas}, \ac{lgnlgn}, and \ac{lgngpd}.  Each distribution is able to capture the salient properties of unimodality and asymmetry typically exhibited by operational loss data.  There are many references that contain candidate loss severity distributions including \citet{Chernobai2007}, \citet{Panjer2006}, and \citet{Peters2015}.

We briefly define both \ac{gnh} and \ac{lsas} and give their distribution function.  Let $Z \sim N(0, 1)$.  Then, \begin{equation*}
X = a + b\cdot A_{g,h}(Z) = a + b \ \Big(\frac{e^{gZ}-1}{g}\Big)e^{hZ^2/2}, \end{equation*}
is said to have a $g$-and-$h$ distribution \citep{Hoaglin1985} where $a \in \R$ is a location parameter and $b > 0$ is a scale parameter.  To ensure monotonicity of the $g$-and-$h$ transformation, $A_{g, h}(\cdot)$, we further assume $g \in \R$ and $h > 0$ for the skewness and elongation parameters, respectively.  As shown by \citet{Degen2007}, these restrictions impose a \ac{rv} right-tail on \ac{gnh} with index $-1/h$.  Thus, we lose the ability to model \ac{subx} and \ac{supx} tail behaviors.

The \ac{lsas} distribution is the result of an exponential transformation of a Sinh-arcSinh \citep{Jones2009} random variable.  It is a four-parameter generalization of the two-parameter lognormal distribution with two additional parameters that allow for more flexible skewness and tailweight.  Let $Z \sim N(0, 1)$.  Then, \begin{align*}
{Y = a + b\cdot A_{\epsilon, \delta}(Z) = a + b \ \sinh\Big\{ \frac{\sinh^{-1}(Z) + \epsilon}{\delta} \Big\}},
\end{align*} is said to have a Sinh-arcSinh distribution with location parameter $a \in \R$, scale parameter $b > 0$, skewness parameter $\ep \in \R$, and tailweight parameter $\de > 0$.  The sinh-arcsinh transformation, $A_{\epsilon, \delta}(\cdot)$, is monotonic over the entire parameter space with a closed-form inverse.  Finally, we say that $X = e^{Y}$ has a \ac{lsas} distribution.

Unlike \ac{gnh}, \ac{lsas} is able to model \ac{rv}, \ac{subx}, and \ac{supx} right-tail behaviors.  Also, the inverse Sinh-arcSinh transformation has a closed-form solution, allowing \ac{mle} to run much faster than for \ac{gnh}, whose inverse transformation requires monotone interpolation of transformed standard normal quantiles for each parameter vector.  The \ac{lsas} likelihood function has better numerical stability than \ac{gnh}, because we can calculate the likelihood on the log of the losses.  Finally, one should be aware that the support for \ac{gnh} is $\mathbb{R}$, which allows for negative losses.  Simulating losses from the unconditional \ac{gnh} using the rejection method may overestimate the probability of a large loss.  Penalized maximum likelihood can be used to reduce the probability of negative losses, but is not used in our analysis.

\subsection{Severity Distribution Selection Criteria} \label{SS:SDSC}

To estimate the candidate severity distributions, we use the \ac{mle} approach which is a well established and well respected method of parameter estimation with many desirable optimality properties formalized by Fisher in 1922 \citep{Aldrich1997}.  Since we employ \ac{mle} to estimate each candidate distribution, a natural choice to compare the estimated distributions is \ac{aic} \citep{Akaike1974}.  The \ac{aic} is defined as
\begin{align}
\label{aic}
AIC &= -2 \ \tilde{\ell}(\widehat{\Theta}; \mathbf{x}, \tau) + 2 \ k, 
\end{align}
where $\tilde{\ell}$ is the log-likelihood function, $\widehat{\Theta}$ is the estimated distribution parameter vector as found via \acs{mle}, and $k$ is the number of estimated parameters in the distribution.  The number of estimated parameters is included in the \acs{aic} to prevent selecting a model that overfits the data.  The \acs{aic} is a relative performance measure, so while it can compare models, it cannot tell if any model is a good fit.  Another weakness of selecting a distribution by \ac{aic} is that operational loss severity data are typically skewed with most of the risk lying in the extreme right tail, but \ac{aic} is a likelihood measure heavily influenced by the central portion of the data's distribution.  Other diagnostics, such as QQ-plots should be consulted for adequacy of tail fit.

Unlike \ac{aic}, the \ac{ad} \citep{Sinclair1990} is a goodness-of-fit test that can be used to eliminate candidate distributions when the data do not follow the estimated distribution with a desired confidence.  The modified Anderson-Darling test statistic is calculated as
\begin{align*}
\widehat{AD} = \frac{n}{2} - 2 \ \sum_{i=1}^n \widetilde{F}(x_{(i)}; \widehat{\Theta}, \tau) - \sum_{i=1}^{n} \Big[ 2 - \frac{2i - 1}{n}\Big]\log\big[1 - \widetilde{F}(x_{(i)}; \widehat{\Theta}, \tau)\big],
\end{align*}
where $n$ is the number of observations, $x_{(i)}$ is the $i^{\text{th}}$ order statistic such that $x_{(1)} \leq x_{(2)} \leq \cdots \leq x_{(n)}$, and $\widetilde{F}(x; \widehat{\Theta}, \tau)$ is the estimated conditional \acs{cdf} for the candidate distribution.

The conditional \ac{cdf} is used to calculate the test statistic, so \ac{ad} can often fail to reject distributions with ``unreasonable" high truncation probability estimates.  This same issue is shared by the Kolmogorov-Smirnov test.  As a result, we incorporate the truncation probability estimate into our severity distribution selection process.  Under the truncation approach, lighter-tailed distributions can mimic heavy-tailed distributions at sufficiently high truncation levels.  For example, Gumbel-type distributions can mimic an inverse power law tail behavior \citep{Perline2005}.  As a result, high truncation probability estimates can signal which distributions may be inappropriate.  From a practical standpoint, extremely high truncation probability estimates, exceeding 90\%, proportionally increase the observed rate of losses which can lead to untenable loss simulations.  We find that high truncation probabilities occur when the \ac{mle} algorithm stops at the boundary of the parameter space, and thus the estimates should not be used.

Since the goal of \ac{or} modeling is to predict the 99.9\% quantile of the firm-wide annual loss distribution from Equations \ref{E:totalrrc} and \ref{E:totalorc}, the final selection process we present in Section \ref{SS:QSAnn} assesses the predictive ability of each estimated severity distribution to predict the 99.9\% quantile of the annual loss distribution for each \ac{orc}.  Under the assumption of weak dependence for the losses from each \ac{orc}, the sum of the forecasted 99.9\% quantiles is a good proxy for \ac{rc}.

Forecasting ability for the estimated distributions is measured by the \acf{qs} \citep{Gneiting2011}.  The \ac{qs} uses a consistent scoring function that is non-negative such that better predictors have smaller scores.  Following \cite{Gneiting2011}, let $L_{T+1, \alpha}^r$ be the predicted $\alpha$-quantile for the annual loss distribution of \ac{orc} $r$.  Then, 
\begin{align*}
L_{T+1, \alpha}^r &= F_{r}^{-1}\Big(\alpha; \lambda^{*}_{r}, \widehat{\bm{\theta}} \Big),
\end{align*}
where $F_{r}^{-1}$ is the quantile function for the annual loss distribution for \ac{orc} $r$, $\alpha$ is a probability between 0 and 1, $\lambda^{*}_{r}$ is the estimated loss frequency parameter for \ac{orc} $r$ (see Section \ref{S:Freq}), and $\widehat{\bm{\theta}}$ is the estimated loss severity parameter vector found via the truncation approach of Section \ref{SS:Trunc}.  Let $\mathbf{L}^r = \{L_t^r\}_{t=1}^{T}$ be a sequence of $T$ observed annual losses from \ac{orc} $r$.  The quantile scoring function is defined as
\begin{align}
\label{E:qs}
S(L_{T+1, \alpha}^r, \mathbf{L}^r) = \frac{1}{T}\sum_{t=1}^T \big(\mathbbm{1}(L_{T+1, \alpha}^r \geq L_t^{r}) - \alpha \big) \big(L_{T+1, \alpha}^r - L_t^{r} \big).
\end{align}
When $\alpha$ is close to 1, the quantile scoring function is asymmetric, penalizing more for underestimation than for overestimation.  This asymmetric feature should be particularly appealing to regulators who want to avoid underestimation of risk.  In Section \ref{SS:QSAnn}, the \ac{qs} is calculated over the top $25\%$ of the annual loss distribution by integrating Equation \ref{E:qs} numerically.  Focusing on a region of the tail instead of a single quantile helps to mitigate complications that can arise if distribution tails cross.

Equation \ref{E:qs} gives an estimate for the expected value of the scoring function under the estimated unconditional annual loss distribution.  Since our data, $\mathbf{L}^{r}$, are the annual sums of observable loss severities over some known threshold, $\tau$, we assume that the probability of having an unobservable annual loss is near zero.

\section{Estimating the Annual Loss Distribution} \label{S:Freq}

From Equation \eqref{E:totalorc}, let $L^{r}_{T+1}$ be a random variable for the annual loss for \ac{orc} $r$ in year $T+1$.  Once an estimated loss severity distribution is selected for an \ac{orc}, we proportionally increase the loss frequency to account for the unobserved losses below the reporting threshold.  Assume the loss frequencies for \ac{orc} $r$ are independent through time and identically distributed as Poisson random variables, $N_{t}^{r}$.  We assume a Poisson frequency distribution throughout this paper.  The rate of observable losses for \ac{orc} $r$ is the mean number of annual loss events from the dataset, $\widehat{\lambda}_r$.  If \ac{orc} $r$'s selected severity distribution is $F_{r}$ with estimated parameters $\widehat{\bm{\theta}}$, then the estimated rate of all annual loss events for \ac{orc} $r$ is
\begin{align*}
\lambda^{*}_r = \dfrac{\widehat{\lambda}_r}{1 - F_{r}\big(\tau; \hat{\bm{\theta}} \big)}.
\end{align*}
Thus, the truncation probability estimate plays an important role in estimating the annual loss distribution.

To numerically derive the estimated distribution of $L^{r}_{T+1}$, we simulate 250,000 annual losses using the estimated frequency and severity distributions and use monotone piecewise cubic hermite interpolation to create a continuous function.  Monotone interpolation is accomplished with the $\texttt{R}$ function $\texttt{pchip()}$ from the $\texttt{signal}$ library.

\section{Simulation Study} \label{S:Sim}

Scaled operational loss data are simulated both above and below a reporting threshold for three unique \ac{orc}'s whose generating processes are given in Table \ref{Tab:Sim}.  To mimic the scenario faced by industry, the \ac{orc} simulations are truncated at a known threshold set in advance.  For each section except Section \ref{SS:cens} where we investigate various truncation points, each \ac{orc}'s truncation point is set at the ${2.5\%}$ quantile of their respective severity distribution.

{\centering
\begin{table}[h!]
$$\textbf{Loss Generating Processes for 3 \ac{orc}'s}$$
\begin{tabular}{ |p{0.8cm}|p{1.9cm}|p{2.2cm}|p{2cm}|p{2.25cm}|p{2cm}|p{2cm}| } 
  \hline
  \multirow{2}{*}{\textbf{ORC}} & \textbf{Truncation} & \textbf{Frequency} & \textbf{Frequency} & \textbf{Severity} & \textbf{Severity} & \textbf{Log 99.9\%} \\
  \hhline{~~~~~~}
   & \textbf{Point} & \textbf{Distribution} & \textbf{Parameters} & \textbf{Distribution} & \textbf{Parameters} & \textbf{Quantile}\\
  \hline
  \hline
  \multirow{3}{*}{1} & \multirow{3}{*}{$\tau = 1.026$} & \multirow{3}{*}{Poisson} & \multirow{3}{*}{$\lambda = 100$} &
   \multirow{3}{*}{Burr} & $\alpha = 0.07$ & \multirow{3}{*}{13.774} \\ 
  \hhline{~~~~~-}
   & & & & & $\gamma = 12$ &  \\ 
  \hhline{~~~~~-}
   & & & & & $\theta = 1.1$ & \\
  \hline
  \hline
  \multirow{4}{*}{2} & \multirow{4}{*}{$\tau = 3.147$} & \multirow{4}{*}{Poisson} & \multirow{4}{*}{$\lambda = 100$} & 
  \multirow{4}{*}{log-SaS} & $a = 1.06$ & \multirow{4}{*}{10.543}\\ 
  \hhline{~~~~~-}
   & & & & & $b = 0.37$ & \\ 
  \hhline{~~~~~-}
   & & & & & $\epsilon = 1.65$ & \\
   \hhline{~~~~~-}
   & & & & & $\delta = 0.97$ & \\
  \hline
  \hline
  \multirow{7}{*}{3} & \multirow{7}{*}{$\tau = 0.923$} & \multirow{7}{*}{Poisson} & \multirow{7}{*}{$\lambda = 100$} & \multirow{1}{*}{$X \sim \beta X_1$} & \multirow{2}{*}{$\beta = 0.33$} & \multirow{7}{*}{13.362}\\ 
  \hhline{~~~~~~}
   & & & & $\quad + (1 - \beta)X_2$ & & \\ 
  \hhline{~~~~--}
   & & & & \multirow{2}{*}{$X_1 \sim \text{LN}$} & $\mu = 0.7$ & \\
  \hhline{~~~~~-}
   & & & & & $\sigma = 0.5$ & \\
  \hhline{~~~~--}
   & & & & \multirow{3}{*}{$X_2 \sim \text{Burr}$} & $\alpha = 0.07$ & \\
  \hhline{~~~~~-}
   & & & & & $\gamma = 12$ & \\
  \hhline{~~~~~-}
   & & & & & $\theta = 1.1$ & \\
  \hline
  \end{tabular}
\centering
\caption[\textbf{Loss Generating Processes for 3 ORC's}: Table of frequency and severity distributions used to simulate operational losses for three ORC's with the natural logarithm of the true 99.9\% quantile of the distribution of the annual losses in the final column]{Table of frequency and severity distributions used to simulate operational losses for three ORC's with the natural logarithm of the true 99.9\% quantile of the distribution of the annual losses in the final column.}
\label{Tab:Sim}
\end{table}
\par}

\subsection{Severity Selection using \ac{aic} and \ac{ad}} \label{SS:aicad}

Random samples of historical losses are simulated 10 times for each \ac{orc}.  Each simulation contains losses for 14 years, $2004, 2005, \dots, 2017$.  For each simulation, the candidate severity distributions are estimated using data for all years $2004, 2005, \dots, T$ for $T = 2013, 2014, \dots 2017$, following the truncation approach of Section \ref{SS:Trunc}.  For each $T$, we select the ``best" severity distribution by first eliminating candidate distributions whose \ac{mle} occur at their parameter space boundary, estimated truncation probability exceeds 0.5, or is rejected by \ac{ad}.  Then, the severity distribution with the lowest \ac{aic} is selected from the remaining candidates and is used to forecast the 99.9\% quantile of the distribution of $L^r_{T+1}$.  This process is repeated over all 10 simulations for each of the 3 \ac{orc}'s.

Figure \ref{Fig:AIC} plots the forecasted $99.9\%$ quantiles for each simulation against the true value, separated by \ac{orc}, on the log scale.  Each plot has the forecasts for years $2014, 2015, \dots, 2018$, for each of the 10 simulations.  Each simulation's forecasts are represented by a dashed or dotted line in the plot, with the true value given as a solid line.  For many simulations, and in particular for \ac{orc} 1 and \ac{orc} 2, we see that the forecasted quantiles can fluctuate wildly year-to-year.  This is the exact problem that practitioners of the \ac{ama} would like to avoid.  

{\centering
\begin{figure}[h!]
\begin{center}
\textbf{Forecasted 99.9\% Quantile when Selecting Severity by \ac{aic}}
\begin{minipage}{0.85\textwidth}
\includegraphics[width=1\linewidth]{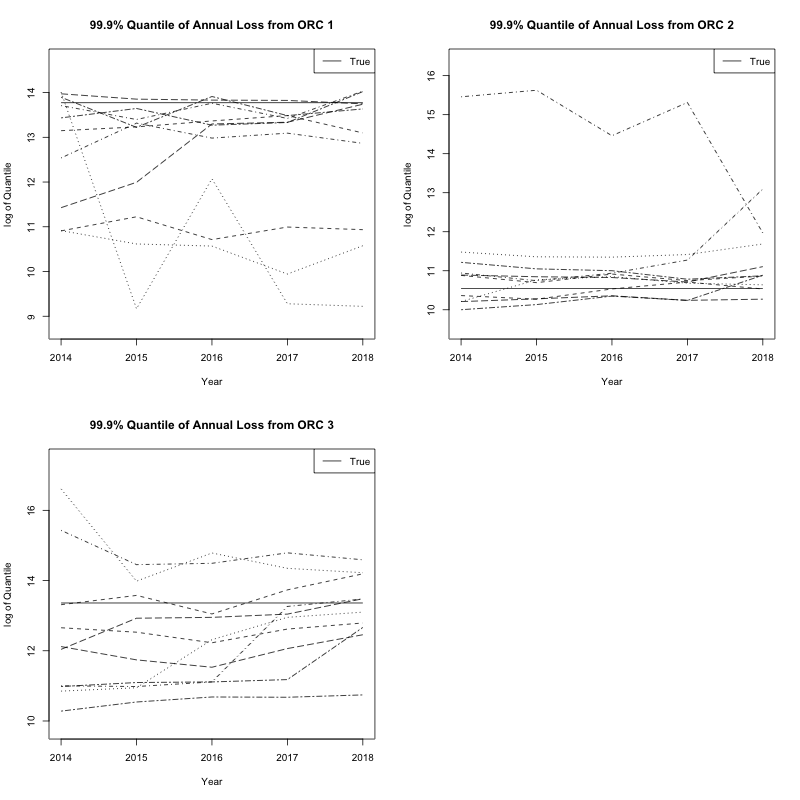}
\caption[\textbf{Forecasted 99.9\% Quantile when Selecting Severity by AIC}: For 10 simulations of losses from each ORC, forecasted 99.9\% quantiles for the distribution of $L^r_{T+1}$ are plotted (dotted or dashed lines) on the log scale against the true value (solid line) for the years $2014, 2015, \dots, 2018$ when selecting the severity distribution by best AIC]{For 10 simulations of losses from each ORC, forecasted 99.9\% quantiles for the distribution of $L^r_{T+1}$ are plotted (dotted or dashed lines) on the log scale against the true value (solid line) for the years $2014, 2015, \dots, 2018$ when selecting the severity distribution by best AIC.}
\label{Fig:AIC}
\end{minipage}
\end{center}
\end{figure}
\par}

Table \ref{Tab:AD} presents the number of times that a severity distribution is accepted as the \ac{orc}'s loss severity distribution when using the \ac{ad} test at the 95\% confidence level.  Since each \ac{orc}, given as rows, is simulated 10 times and each simulation has 5 years of forecasts, there are 50 accept/reject decisions from the \ac{ad} test for each severity candidate.  A candidate distribution that is never rejected has 50 accept decisions.  The low power of the \ac{ad} test is particularly bad for \ac{orc}'s 1 and 2, whose underlying generating process is a single parametric distribution.  As mentioned in Section \ref{SS:SDSC}, this is in large part due to high truncation probability estimates enabling thinner-tailed distributions to mimic fatter-tailed behavior.

{\centering
\begin{table}[h!] 
\begin{center}
\textbf{Modified Anderson-Darling Hypothesis Tests}
\includegraphics[width=0.7\linewidth]{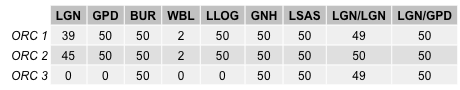}
\caption[\textbf{Modified Anderson-Darling Hypothesis Tests}: 50 hypothesis tests using the modified Anderson-Darling test at the 95\% significance level are performed for each candidate severity distribution, and the number of accept decisions when losses are simulated for each ORC are given in the rows]{50 hypothesis tests using the modified Anderson-Darling test at the 95\% significance level are performed for each candidate severity distribution, and the number of accept decisions when losses are simulated for each \ac{orc} are given in the rows.}
\label{Tab:AD}
\end{center}
\end{table}
\par}

\subsection{$g$-and-$h$ versus log-SaS} \label{SS:ghsas}

For a randomly chosen simulation of each \ac{orc} from Section \ref{SS:aicad}, we compare the impact on the 99.9\% quantile of the estimated annual loss distribution when selecting either \ac{lsas} or \ac{gnh} as the severity distribution.  When either model is appropriate, such as in \ac{orc} 1 or \ac{orc} 2,  both distributions produce similar estimates over time. Figure \ref{Fig:GHSAS} plots the $99.9\%$ quantiles of the estimated annual loss distribution for each \ac{orc} when using \ac{lsas} and \ac{gnh} as the severity distribution.  Since \ac{lsas} offers a number of advantages over \ac{gnh} outlined in Section \ref{SS:Dist}, \ac{lsas} is a viable candidate for modeling operational loss severities.

\ac{orc} 3 is a scenario where neither \ac{lsas} nor \ac{gnh} is an appropriate severity distribution.  The overestimation from \ac{gnh} can be explained by looking at the \ac{mle} parameters.  For each year, the estimated \ac{gnh} parameters produce a negatively skewed distribution with a probability of negative loss greater than $40\%$.  When simulating loss severities to derive the annual loss distribution for \ac{orc} 3, the rejection method overestimates the probability for large losses.  The underestimation of the \ac{lsas} distribution shows how the \ac{mle} approach, which is heavily influenced by the central portion of the data, may impact the estimated tail behavior of the estimated loss severity distribution.

{\centering\begin{figure}[h!]
\begin{center}
\textbf{Forecasted 99.9\% Quantile when Selecting \ac{gnh} or \ac{lsas} Severity}
\begin{minipage}{0.9\textwidth}
\includegraphics[width=1\linewidth]{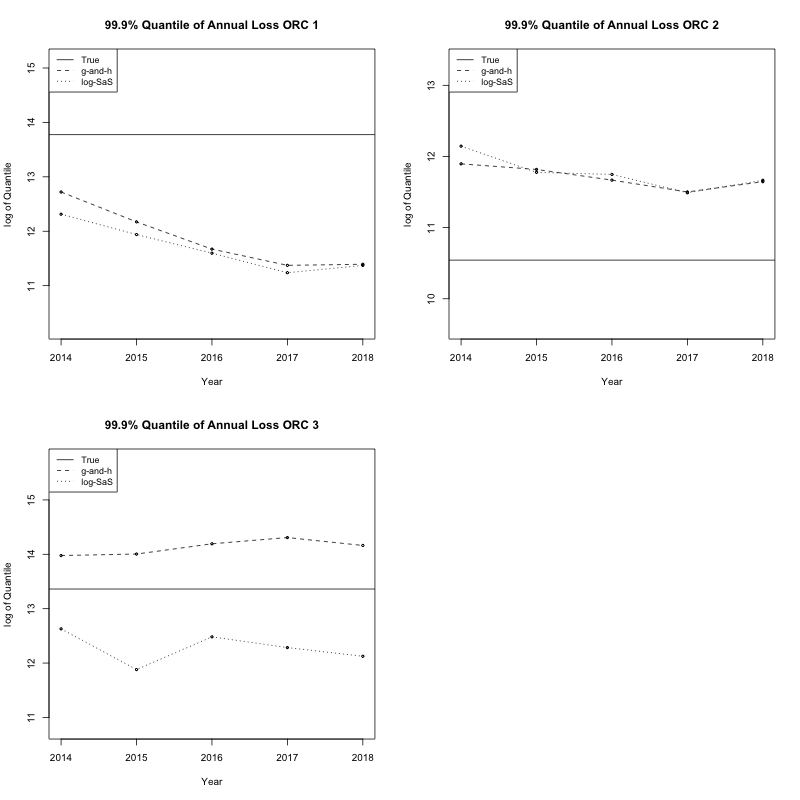}
\caption[\textbf{Forecasted 99.9\% Quantile when Selecting GNH or LSAS Severity}: Forecasted 99.9\% quantiles for the distribution of $L^r_{T+1}$ are plotted on the log scale against the true value (solid line) when selecting the GNH (dashed line) and the LSAS (dotted line) for the years $2014, 2015, \dots, 2018$]{Forecasted 99.9\% quantiles for the distribution of $L^r_{T+1}$ are plotted on the log scale against the true value (solid line) when selecting the FNH (dashed line) and the LSAS (dotted line) for the years $2014, 2015, \dots, 2018$.}
\label{Fig:GHSAS}
\end{minipage}
\end{center}
\end{figure}
\par}

\subsection{Censored Data} \label{SS:cens} 

This section explores improvements in the truncation probability estimates when loss frequency data are collected for operational losses below the reporting threshold.  We simulate 2000 samples, where each sample is comprised of 14 years worth of losses using the \ac{orc} 1 parameters given in Table \ref{Tab:Sim}.  Let the true loss severity parameter vector be represented by $\bm{\theta}_0$.  We truncate the data at the true $2.5\%, 5\%, 10\%$, and $20\%$ quantiles of the underlying Burr distribution, use the Burr distribution as our only candidate family, and estimate the truncation and censoring probabilities.

Table \ref{Tab:TrPr} gives the means and standard deviations of the truncation and censoring probability estimates at each truncation point.  The final column, Simulations, indicates the number of simulations where the \ac{mle} algorithm was able to converge for both the truncation and censoring approaches at each truncation point.  Only these simulations are included when calculating the mean and standard deviations.  Collecting the frequency of losses below the reporting threshold significantly reduces both the bias and variability of the truncation probability estimate.

{\centering
\begin{table}[h!]
$$ \textbf{Truncation/Censoring Probability Estimates} $$
\begin{tabular}{ |p{1.5cm}|p{1.5cm}|p{2cm}|p{1.5cm}|p{2cm}|p{2cm}|} 
  \hline
  \multirow{2}{*}{$F(\tau; \bm{\theta}_0)$} & \multirow{2}{*}{$\overline{F(\tau; \widehat{\bm{\theta}})}$} & 
  \multirow{2}{*}{SD\Big({$F(\tau; \widehat{\bm{\theta}})$}\Big)} & \multirow{2}{*}{$\overline{F(\tau; \widehat{\bm{\theta}}_c)}$} 
  & \multirow{2}{*}{SD\Big({$F(\tau; \widehat{\bm{\theta}}_c)$}\Big)} & \multirow{2}{*}{Simulations} \\
  \hhline{~~~~~}
   & & & & & \\
  \hline
  \hline
   0.025 & 0.027 & 0.016 & 0.025 & 0.004 & 1967 \\
  \hline
  \hline
  0.05 & 0.072 & 0.085 & 0.05 & 0.006 & 1779 \\ 
  \hline
  \hline
  0.1 & 0.206 & 0.234 & 0.1 & 0.008 & 1263 \\ 
  \hline
  \hline
  0.2 & 0.398 & 0.287 & 0.2 & 0.011 & 860 \\
  \hline
   \end{tabular}
\centering
\caption[\textbf{Truncation/Censoring Probability Estimates}: From 2000 simulations of losses for ORC 1 in Table \ref{Tab:Sim}, we show (from left-to-right) the true truncation/censoring probability, the mean and standard deviation of the truncation probability estimates, the mean and standard deviation of the censoring probability estimates, and the number of simulations where the MLE algorithms converged for both approaches]{From 2000 simulations of losses for ORC 1 in Table \ref{Tab:Sim}, we show (from left-to-right) the true truncation/censoring probability, the mean and standard deviation of the truncation probability estimates, the mean and standard deviation of the censoring probability estimates, and the number of simulations where the MLE algorithms converged for both approaches.}
\label{Tab:TrPr}
\end{table}
\par}

\subsection{Quantile Score of Annual Losses} \label{SS:QSAnn}

The \ac{qs} can be used on annual loss data to select a model that produces the most accurate probabilistic forecast of the tail.  The following process is performed for each \ac{orc}, respectively:
\begin{enumerate}
\itemsep0em 
  \item{50 years of losses are simulated given the parameters in Table \ref{Tab:Sim}.}
  \item{Losses are truncated at the \ac{orc} truncation point.}
  \item{Loss frequency and severity distribution parameters are estimated from the observable loss data using the truncation approach.}
  \item{Candidate distributions are eliminated if the \ac{mle} parameters occur at a boundary or if the truncation probability is 0.5 or higher.}
  \item{For the remaining severity distributions, the annual loss distribution is estimated via simulation.}
  \item{The observable losses from Step 2 are aggregated by year and treated as our annual loss data.}
  \item{The \ac{qs} is calculated for the upper 25\% of the tail using the observed annual losses from Step 6 and the estimated annual loss distributions from Step 5.}
  \item{The \ac{qs}'s are ranked by distribution from best (1) to worst (9), where (9) represents the distribution is eliminated in Step 4.}
  \item{Steps 1 thru 8 are repeated 100 times to capture sampling variability.}
  \item{Boxplots of the ranks for each distribution's \ac{qs} are presented in Figure \ref{Fig:09QSRank}.}
\end{enumerate}
  
{\centering
\begin{figure}[h!]
\begin{center}
\textbf{Candidate Severity Distribution \ac{qs} Ranks for Annual Losses}
\begin{minipage}{0.9\textwidth}
\includegraphics[width=1\linewidth]{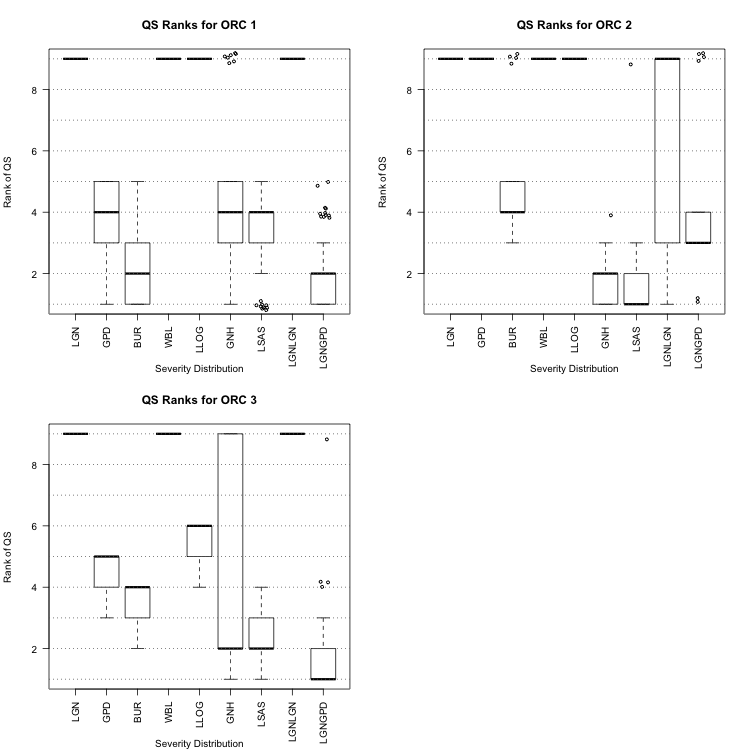}
\caption[\textbf{Candidate Severity Distribution QS Ranks for Annual Losses}: For each ORC, boxplots of the ranks by QS for 100 simulations of 50 years of losses for each ORC, where the worst rank, 9, occurs either when an estimated severity distribution's parameter occur at a boundary or the truncation probability exceeds 0.5]{For each \ac{orc}, boxplots of the ranks by \ac{qs} for 100 simulations of 50 years of losses for each \ac{orc}, where the worst rank, 9, occurs either when an estimated severity distribution's parameter occur at a boundary or the truncation probability exceeds 0.5.}
\label{Fig:09QSRank}
\end{minipage}
\end{center}
\end{figure}
\par}

Figure \ref{Fig:09QSRank} shows the distribution of 100 \ac{qs} ranks from each candidate severity distribution for the samples outlined above.  For \ac{orc} 1, the true underlying severity distribution is \ac{bur} with \ac{rv} tail behavior.  As a result, we see \ac{subx} distributions eliminated and given ranks 9 for each of the 100 samples.  \ac{orc} 3 is generated by a mixture distribution.  The closest severity candidate is \ac{lgngpd}, which we see as performing the best.  \ac{lsas} and \ac{gnh} perform second best, but one should note that \ac{gnh} is often eliminated from consideration due to boundary conditions or high truncation probabilities leading to a large proportion of rank 9.  Finally, we note that the true underlying tail comes from \ac{bur}, whose \ac{qs}'s tend to fall behind \ac{lgngpd}, \ac{lsas}, and \ac{gnh}.  This is due to the \ac{mle} approach for estimating loss severity parameters.  Since the \ac{bur} parameters are estimated using all of the loss severities, the parameters are influenced by the body of the distribution at the expense of the tail.

Figure \ref{Fig:09QSRank} also showcases the \ac{lsas} distribution.  Many outliers produced by \ac{gnh} for \ac{orc}'s 1 and 3 showcase the flexibility of the \ac{lsas} distribution and the stability of its likelihood function.

\section{Conclusion} \label{S:Conc}

In conclusion, we feel that the \ac{qs} and truncation probability estimate should be used in conjunction with likelihood and goodness-of-fit measures when selecting a severity distribution.  While we used \ac{mle} to estimate each candidate distribution, one may want to explore using the \ac{qs} as an objective function to minimize for parameter estimation.

We prefer \ac{lsas} to \ac{gnh} as a loss severity candidate due to its ability to model various tail behaviors and ease of numerical derivation.  However, both distributions can be added as candidates.  The \ac{gnh} log-likelihood function is numerically more difficult to work with than the \ac{lsas} log-likelihood function, with more instances of non-convergence of numerical maximum likelihood for \ac{gnh}.  Including both \ac{gnh} and \ac{lsas} as candidate distribution may preserve at least one of the four-parameter candidates when the numerical maximum likelihood is unable to converge for \ac{gnh}.  Future research should investigate penalized maximum likelihood approaches for \ac{gnh} to minimize the probability of negative losses.

The \ac{qs} performs remarkably well on annual loss data with only 50 years of data.  While banks may have to wait 20-30 more years to have that much data available, the performance of the \ac{qs} is very good with only 50 data points and should be used when selecting the loss severity.

\section*{Acknowledgements and Declaration of Interest}

The authors would like to acknowledge the Scotiabank Cybersecurity and Risk Analytics Initiative for funding this research.

\bibliographystyle{chicago} 
\bibliography{/Users/danielhadley/Documents/BibFiles/bib}

\newpage

\appendix

\section{Maximum Likelihood Estimates for ORC 3 when Comparing \ac{gnh} to \ac{lsas}} \label{S:MLE}
\begin{center}
\begin{figure}[h!]
\begin{center}
\includegraphics[width=1\linewidth]{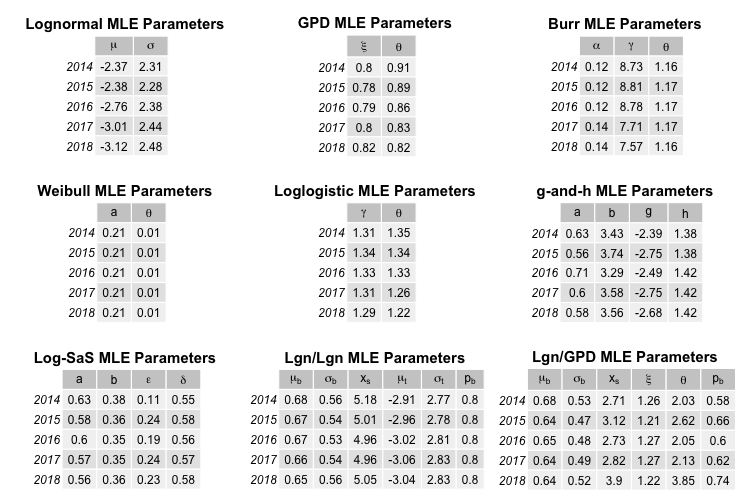}
\caption{The \ac{mle} from the simulated data \ac{orc} 3 for each candidate severity distribution for the years {2014~--~2018}}
\label{Fig:Burr}
\end{center}
\end{figure}
\end{center}

\newpage

\section{Loss Severity Distributions} \label{S:LSD}
When describing the right-tail behavior for the following loss severity distributions, the following notations are used: \acf{supx}, \acf{subx}, and \acf{rv}.

Let $\Phi(\cdot)$ and $\phi(\cdot)$ represent the standard normal \ac{cdf} and \ac{pdf}, respectively.  Let \begin{equation*}A^{-1}_{g,h}\Bigg( \frac{x-a}{b} \Bigg)\end{equation*} be the inverse $g$-and-$h$ transformation function with location parameter $a$ and scale parameter $b$, and let \begin{equation*}A_{\epsilon, \delta}^{-1}\Bigg(\frac{\log(x) - a}{b}\Bigg)\end{equation*} be the inverse log-SaS transformation function with location parameter $a$ and scale parameter $b$.

For the spliced distributions, $p_b$ is the proportion of the sample data that falls in the body of the sample, $x_s$ is the splicing point,
\begin{align*}
D_1({\Theta}, \tau) &= F_{body}(x_s; {\Theta}_b) - (1-p_b)F_{body}(\tau; {\Theta}_b); \\
\\D_2({\Theta}, \tau) &= \frac{1-p_b}{D_1({\Theta}, \tau)} \ \frac{F_{body}(x_s; {\Theta}_b) - F_{body}(\tau; {\Theta}_b)}{1 - F_{tail}(x_s; {\Theta}_u)},
\end{align*}
where $F_{body}$ is the lognormal distribution, and $F_{tail}$ is lognormal for the \ac{lgnlgn} distribution and generalized Pareto for the \ac{lgngpd} distribution.


\begin{table}[h!]
$$\textbf{Candidate Severity Distribution Parameterizations}$$
\begin{tabular}{|p{2cm}|p{6cm}|p{2.4cm}|p{4.1cm}| } 
  \hline
  \textbf{Distribution} & \textbf{$F(x; \Theta)$} & \textbf{Support} & \textbf{Right Tail} \\
  \hline
  \hline
  Lognormal & $\Phi\Big[ \frac{\log(x)-\mu}{\sigma}\Big]$ & $x \in \mathbb{R}$ & \acs{subx} \\ 
  \hline
  \hline
  \multirow{2}{*}{Generalized} & \multirow{3}{*}{$1 - \Big( 1 + \frac{\xi}{\theta}(x - u) \Big)^{-1/\xi}$} & $x > u$ & $\xi > 0 \implies$ \acs{rv} \\
  \hhline{~~--}
   \multirow{2}{*}{Pareto} & & $x > u$ & $\xi = 0 \implies$ Exponential \\
   \hhline{~~--}
   & & $u < x < u - \theta/\xi$ & $\xi < 0 \implies$ Bounded above \\
  \hline
  \hline
  Burr & $1 - \Big[ 1 + \big(\frac{x}{\theta}\big)^{\gamma} \Big]^{-\alpha}$ & $x > 0$ & \acs{rv} \\
  \hline
  \hline
  \multirow{2}{*}{Weibull} & \multirow{2}{*}{$1 - \exp\Big\{ -\big[\frac{x}{\theta}\big]^a \Big\}$} &\multirow{2}{*}{$x > 0$} & $a < 1 \implies$ \acs{subx} \\
   \hhline{~~~-}
    & & & $a > 1 \implies$ \acs{supx} \\
    \hline
    \hline
   Loglogistic & $\Big[1 + \big( x / \theta \big)^{-\gamma}\Big]^{-1}$ & $x > 0$ & \acs{rv} \\
    \hline
    \hline
  $g$-and-$h$ & $\Phi\Big[ A^{-1}_{g,h}\Big( \frac{x-a}{b} \Big) \Big]$ & $x \in \mathbb{R}$ & $h > 0 \implies$ \acs{rv}\\
    \hline
    \hline
   \multirow{2}{*}{Log} & \multirow{3}{*}{$\Phi\Big[ A_{\epsilon, \delta}^{-1}\Big(\frac{\log(x) - a}{b}\Big) \Big]$} & \multirow{3}{*}{$x > 0$} & $\delta \leq 0.5 \implies$ \acs{rv} \\
   \hhline{~~~-}
   \multirow{2}{*}{sinh-arcsinh} & & & $0.5 < \delta < 1 \implies$ \acs{subx} \\
    \hhline{~~~-}
     & & & $\delta > 1 \implies$ \acs{supx} \\
    \hline
    \hline
    \multirow{2}{*}{Lognormal} & \multirow{3}{*}{$\dfrac{p_b}{D_1({\Theta}, \tau)}\Phi\Big[ \frac{\log(x)-\mu_b}{\sigma_b}\Big]$} & \multirow{3}{*}{$0 < x \leq x_s$} & \multirow{8}{*}{\acs{subx}}\\
   \hhline{~~~~}
   \multirow{2}{*}{Body} & & & \\
   \hhline{~~~~}
   \multirow{2}{*}{Spliced} & & & \\
   \hhline{~--~}
    & \multirow{1}{*}{$\dfrac{p_b\Phi\Big[ \frac{\log(x)-\mu_b}{\sigma_b}\Big]}{D_1({\Theta}, \tau)} + D_2({\Theta}, \tau) \Phi\Big[ \frac{\log(x)-\mu_t}{\sigma_t}\Big]$} & \multirow{3}{*}{$x_s < x$} & \\
   \hhline{~~~~}
   \multirow{2}{*}{Lognormal} & & & \\
   \hhline{~~~~}
   \multirow{2}{*}{Tail} & \multirow{4}{*}{$\qquad - D_2({\Theta}, \tau)\Phi\Big[ \frac{\log(x_s)-\mu_t}{\sigma_t}\Big]$} & & \\
   \hhline{~~~~}
    & & & \\
    \hhline{~~~~}
    & & & \\
  \hline
  \hline
    \multirow{2}{*}{Lognormal} & \multirow{3}{*}{$\dfrac{p_b}{D_1({\Theta}, \tau)}\Phi\Big[ \frac{\log(x)-\mu_b}{\sigma_b}\Big]$} & \multirow{3}{*}{$0 < x \leq x_s$} & \multirow{8}{*}{\acs{rv}}\\
   \hhline{~~~~}
   \multirow{2}{*}{Body} & & & \\
   \hhline{~~~~}
   \multirow{2}{*}{Spliced} & & & \\
   \hhline{~--~}
   & & & \\
   \hhline{~~~~}
   \multirow{2}{*}{Generalized} & \multirow{1}{*}{$\dfrac{p_b\Phi\Big[ \frac{\log(x)-\mu_b}{\sigma_b}\Big]}{D_1({\Theta}, \tau)} + D_2({\Theta}, \tau)$} & \multirow{3}{*}{$x_s < x$} & \\
   \hhline{~~~~}
   \multirow{2}{*}{Pareto Tail} & \multirow{4}{*}{$\qquad - D_2({\Theta}, \tau)\Big\{1 + \frac{\xi}{\hat{\theta}}(x - x_s)\Big\}^{-1/\xi}$} & & \\
   \hhline{~~~~}
    & & & \\
    \hhline{~~~~}
    & & & \\
  \hline
\end{tabular}
\centering
\caption{Table of candidate severity distributions, their parameterizations, support, and behavior of the right-tail of the density function, where \acf{rv} right tail if its density decreases to 0 at the rate $x^{-b}$ with $b > 1$, \acf{subx} if its 
  density decreases to 0 slower than $e^{-x}$, but faster than \acs{rv}, or \acf{supx} if its density decreases to 0 faster than $e^{-x}$ \mbox{\citep{Foss2013}}.}
\label{Tab:Dists}
\end{table}
\end{document}